\documentclass[aps,prx, superscriptaddress, showpacs,twocolumn,nofootinbib,longbibliography,nobalancelastpage]{revtex4-2}

\usepackage{xcolor}

\usepackage[normalem]{ulem}

\usepackage{hyperref}
\hypersetup{colorlinks=true, citecolor=blue, urlcolor=blue, linkcolor=blue}

\usepackage[normalem]{ulem}
\usepackage{cancel}

\usepackage{graphicx, color}
\usepackage{dcolumn}
\usepackage{bm}
\usepackage{amsmath}
\usepackage{amssymb}
\usepackage{verbatim}
\usepackage{appendix}
\usepackage{float}
\usepackage{color, soul}
\usepackage{dsfont}
\usepackage{mathtools}
\usepackage{blkarray, bigstrut}

\definecolor{brown}{rgb}{0.59, 0.29, 0.0}
\definecolor{darkcyan}{rgb}{0.0, 0.55, 0.55}
\definecolor{darkspringgreen}{rgb}{0.09, 0.45, 0.27}
	\definecolor{darkviolet}{rgb}{0.58, 0.0, 0.83}

\begin{document}

\title{Using parity-nonconserving spin-spin coupling to measure the Tl nuclear anapole moment in a TlF molecular beam}

\author{John W. Blanchard}
\affiliation{Helmholtz-Institut Mainz, GSI Helmholtzzentrum f{\"u}r Schwerionenforschung GmbH, 55128 Mainz, Germany}
\affiliation{Quantum Technology Center, University of Maryland, College Park, MD 20742, USA}

\author{Dmitry Budker}
\affiliation{Helmholtz-Institut Mainz, GSI Helmholtzzentrum f{\"u}r Schwerionenforschung GmbH, 55128 Mainz, Germany}
\affiliation{Johannes Gutenberg-Universit{\"a}t Mainz, 55128 Mainz, Germany}
\affiliation{Department of Physics, University of California, Berkeley, CA 94720-7300 USA}

\author{David DeMille}
\affiliation{Department of Physics, University of Chicago, Chicago, IL 60637 USA}
\affiliation{Physics Division, Argonne National Laboratory, Lemont, IL 60439 USA}

\author{Mikhail G. Kozlov}
\affiliation{Petersburg Nuclear Physics Institute of NRC “Kurchatov Institute”, Gatchina 188300, Russia}
\affiliation{St. Petersburg Electrotechnical University ``LETI'', St. Petersburg 197376 Russia}

\author{Leonid V. Skripnikov}
\affiliation{Petersburg Nuclear Physics Institute of NRC “Kurchatov Institute”, Gatchina 188300, Russia}
\affiliation{Saint Petersburg State University, 7/9 Universitetskaya nab., St. Petersburg, 199034 Russia}

\date{10.11.2022}

\begin{abstract}
An experiment utilizing a TlF molecular beam is being developed by the CeNTREX collaboration to search for hadronic interactions that violate both time-reversal (T) and parity (P) invariance. Here we propose to use the same beam to look for a T-invariance conserving but P-nonconserving (PNC) effect induced by the anapole moment of the Tl nucleus, via a vector coupling of the two nuclear spins in TlF.  
To measure the nuclear anapole moment, the dc electric and magnetic fields in CeNTREX are replaced by rf fields resonant with a nuclear spin flip transition. We adapt the relativistic coupled cluster method in a combination with relativistic density functional theory for the calculation of the molecular PNC spin-spin vector coupling constant that links the experimental signal with the anapole moment. The value of the P-conserving spin-spin coupling constant calculated within the same approach is found to be in good agreement with available experimental data.
\end{abstract}


\maketitle

\section{Introduction}


Spatial parity (P) symmetry is violated in the weak interactions. Several atomic experiments have been performed to study this phenomenon as reviewed in Refs.\,\cite{Bouchiat:2011,GFreview,Safronova:18,Khr91}. The nuclear spin-independent P-odd effect that arises mainly due to the exchange of Z$^0$-bosons between electrons and the nucleus has been measured several times using atoms and is well-established~\cite{Safronova:18}. However, the much smaller nuclear spin-dependent P-odd effects --- dominated, for atoms with large atomic number $Z$, by interaction of electrons with the nuclear anapole moment---have been observed in atoms only once, with $14\%$ uncertainty~\cite{Wood:1997}.  

Molecules are promising systems to study parity nonconserving (PNC) effects~\cite{SF78,Labz78,KL95,Khr91}, as they have close levels of opposite parity. The small energy interval leads to the mixing of these states by electroweak interactions being amplified. Even so, molecular PNC remains as yet undetected. Various types of molecular experiments have been proposed. A Stark-interference approach was considered and implemented in Refs.\,\cite{DeMille:2008,AACD18}. It also has been suggested to use the optical rotation technique for diatomic molecules~\cite{SF78,KL95,Geddes:18}, or to employ enantiomers of chiral molecules~\cite{Gajzago:1974,Letokhov:1975,Kompanets:1976,Harris:1980,Gorshkov1982p4,Barra1996parity,Bauder:1997,Ville:2005,Laubender:2006,Bast:2006,Weijo:2007,Nahrwold:2014,Eills:2017}. Following the proposal in Ref.\,\cite{Blanchard:2020}, we consider here a new type of experiment to probe the nuclear spin-dependent PNC effect, using a non-chiral molecule. It aims specifically to measure the PNC contribution to the indirect nuclear spin-spin coupling in a diatomic molecule. 

To introduce this effect, let us first consider a diatomic molecule with closed electronic shells, with nuclei carrying nonzero nuclear spins $\mathbf{I}$ and $\mathbf{S}$, respectively. In general, the effective Hamiltonian of the indirect interaction of these spins can be written in the following form (we use units with $\hbar=1$ throughout the paper):
\begin{equation}
{H_{J}}=2\pi {I_{i} J_{ik} S_{k}}.
\label{tensor}
\end{equation}
Here $J_{ik}$ is a reducible rank-2 tensor. The rank-1 contribution  is given by
\begin{equation}
{H^{(1)}}=2\pi\mathbf{J}^{(1)} \cdot ( \mathbf{I}\times \mathbf{S} ),
\label{rank1_a}
\end{equation}
where 
\begin{equation}
J^{(1)}_i=\frac{1}{2} \epsilon_{ijk}J^{(1)}_{jk},
\label{rank1}
\end{equation}
$\epsilon_{ijk}$ is the Levi-Civita tensor.
We limit ourselves to the discussion of time-reversal-invariant interactions, so the only option for the ${\bf J}^{(1)}$ vector is to be directed along the molecular axis $\bm \lambda$ (a unit vector directed from the heavier nucleus to the lighter one). This is a polar vector, so the resultant interaction \eqref{rank1_a} is parity nonconserving.  Hence we write 
\begin{equation}
{H_{\mathrm{PNC}}} \equiv {H^{(1)}}= 2\pi J^{(1),\mathrm{PNC}} \bm{\lambda} \cdot ( \mathbf{I}\times \mathbf{S} ),
\label{Jpnc_a}
\end{equation}
where we write the additional superscript (PNC) to emphasize the parity-violating nature of this term.

Here we choose the TlF molecule as an example system to discuss possible measurement of the  PNC $J$-coupling effect. Currently, TlF is used in the Cold molecule Nuclear Time-Reversal EXperiment (CeNTREX)~\cite{Norrgard:2017,CeNTREX21}. CeNTREX aims to measure the effect induced by the T,P-violating interaction of the nuclear Schiff moment with electrons, with the goal of increasing the sensitivity by three orders of magnitude with respect to the best previous experiment on TlF~\cite{Cho:1991}. In CeNTREX, the measurement scheme is to use an external electric field to orient the molecules along or against the Tl nuclear spin, and then measure the spin flip energy. The idea of the present proposal is to use similar methods to measure $J^{(1),\mathrm{PNC}}$, which is induced mainly by the P-odd, T-even nuclear anapole moment of Tl. Below we outline a possible scheme of the experiment and then make estimates of the expected signals and associated sensitivity to PNC. These estimates are based on a precise \textit{ab-initio} study of the electronic structure of TlF, with an accurate treatment of the relativistic and electron-correlation effects, which is described here. The expected PNC effect is expressed in terms of the dimensionless constant $g$ characterizing the nuclear anapole moment of the $^{205}$Tl nucleus (see Sec.\,\ref{Sec:EST}). According to Refs.\,\cite{Flambaum:84,GFreview}, $g(^{205}{\rm Tl})\approx0.5$.


\section{The system}

The $^{205}$TlF molecule has closed electronic shells in its ground electronic state (X $^1\Sigma^+$). The ground-state levels can be described in terms of the rotational quantum numbers $J,M_J$ and the spins $I=1/2, M_I$ and $S=1/2, M_S$ of the $^{205}$Tl and $^{19}$F nuclei, respectively. The rotational Hamiltonian is $H_{\mathrm{rot}}= B\mathbf{J}^2$, and the spins and rotation are coupled via hyperfine interactions, described by the Hamiltonian~\cite{Boeckh1964,Wilkening:1984,CeNTREX21}:
\begin{equation}
    H_{\mathrm{HFS}} = c_1\mathbf{I}\cdot\mathbf{J}+c_2\mathbf{S}\cdot\mathbf{J}+c_3 T^{(2)}(\mathbf{C})\cdot T^{(2)}(\mathbf{I},\mathbf{S})+c_4\mathbf{I}\cdot\mathbf{S}.
    \label{H_HFS}
\end{equation}
In TlF, the hyperfine couplings $c_i$ are of order kHz while the rotational constant is $B \approx 6.67$ GHz.  
Here the third term is the scalar product of two rank-2 tensors: one constructed from two copies of the modified spherical harmonics $\bm{C}$ and the other from $\mathbf{I}$ and $\mathbf{S}$~\cite{brown2003}. 
The second-order hyperfine interaction contributes to the third and the fourth terms of the Hamiltonian \eqref{H_HFS}, while the direct dipole-dipole interaction between nuclear magnetic moments contributes only to the third one.
An external electric field, $\bm{\mathcal{E}}$, leads to the Stark Hamiltonian $H_{\mathrm{St}} = -\mathbf{d}\cdot \bm{\mathcal{E}} = d \bm{\lambda} \cdot \bm{\mathcal{E}}$, where $d = 4.228$ D is the TlF electric dipole moment \cite{Dijkerman1972TlFStark}. 

\begin{figure}[htb]
    \centering
    \includegraphics[width=\columnwidth]{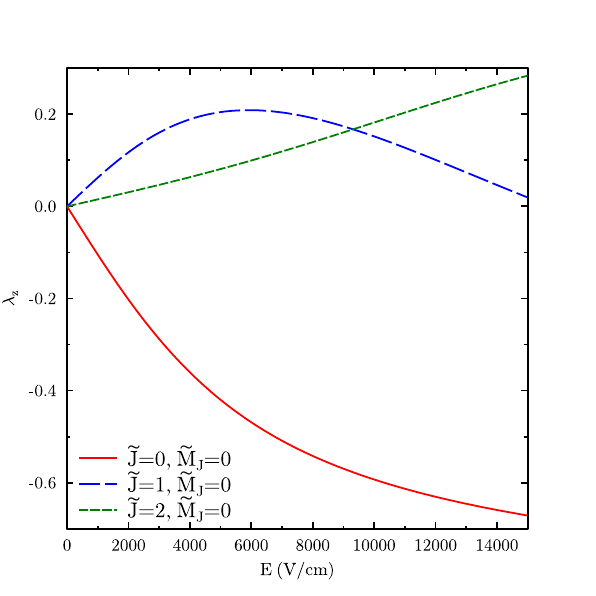}
    \caption{Polarization $\langle \lambda_z\rangle$ as a function of the electric field, for the lowest $J$ states with $m_J=0$. The curves are calculated by numerical diagonalization of the rotational + Stark Hamiltonian.}
    \label{fig:n_z}
\end{figure}

For now, we restrict our discussion to TlF levels in the lowest rotational state, $J=0$. Applying an electric field, $\bm{\mathcal{E}} = \mathcal{E}\hat{\bm{z}}$, will electrically polarize the molecule such that the expectation value of the molecular axis direction, $\langle \bm{\lambda} \rangle = \langle \lambda_z \rangle \hat{\bm{z}}$, is nonzero (see Fig.~\ref{fig:n_z}). 
The polarization arises, to first order, from a mixture of $J=1$ odd-parity states into the $J=0$ even-parity states. For sufficiently small values of $\mathcal{E}$, such that $\mathcal{E} \ll \mathcal{E}_p \equiv B/d$, from perturbation theory it is found that the dimensionless polarization $\langle \lambda_z \rangle$ is linear in $\mathcal{E}$.  We define 
\begin{equation}
    \langle\lambda_z \rangle(\mathcal{E}) \approx \lambda_{z1} (\mathcal{E}/\mathcal{E}_p),
\end{equation}
where $\lambda_{z1} = -1/(2\sqrt{3})$ from angular factors. In TlF, $\mathcal{E}_p \approx 3.12$\,kV/cm~\cite{Dijkerman1972TlFStark}. 

The $|J=0,M_J=0\rangle$ state in TlF has four spin sublevels. In a weak $\mathcal{E}$-field such that $J=0$ remains an approximately good quantum number, the $J$-dependent terms in Eq.\,\eqref{H_HFS}, proportional to $c_1$, $c_2$, and $c_3$, vanish. This can be seen from the explicit expression of matrix elements which are diagonal in $J$ for the case of $J=0$ (see also an explicit form of the matrix element of the term proportional to $c_3$ in Ref.\,\cite{Wilkening:1984}). The remaining (second-order) hyperfine interaction term, $c_4\mathbf{I}\cdot\mathbf{S}$, couples the spin sublevels into a total spin triplet $\mathrm{T}$ and a singlet $\mathrm{S}$. We denote these states---including their $\bm{z}$-projection quantum numbers, $M$, as subscripts---as $|T_0\rangle$, $|T_{\pm 1}\rangle$, and $|S_0\rangle$. 

Now consider the two-level system $|\mathrm{S_0}\rangle$ and $|\mathrm{T_0}\rangle$, with nonzero splitting $\Delta$ resulting from the hyperfine spin-spin interaction.\footnote{A small modification to $\Delta$ from the Stark interaction is discussed later.}
These levels can be coupled to each other by an external magnetic field, $\bm{\mathcal{B}}= \mathcal{B}\hat{\bm{z}}$. In the presence of $\bm{\mathcal{E}}$, parity mixing due to the Stark interaction induces a nonzero value of $\langle \lambda_z \rangle$, so these states will then also be coupled by the PNC interaction of Eq.\,\eqref{Jpnc_a}. 

The effective Hamiltonian for the system is then
\begin{equation}\label{Hamiltonian}
 H =\,\,
  \begin{blockarray}{*{2}{c}}
    \begin{block}{*{2}{>{$\footnotesize}c<{$}}}
      $\left\langle \mathrm{S_0}\right|$ &  $\left\langle \mathrm{T_0}\right|$  \\
    \end{block}
    \begin{block}{(*{2}{c})}
      0  & 0   \\
     0 & \Delta   \\
    \end{block}\\
  \end{blockarray}
  \,\,+H_\mathrm{Z}+H_\mathrm{PNC}\,.
  \vspace{-12pt}
\end{equation}
Here, the Zeeman Hamiltonian, $H_\mathrm{Z}$ may be written as
\begin{align}
  H_{\mathrm{Z}} &= -\bm{\mathcal{B}} \cdot \left(\gamma_I \mathbf{I} +\gamma_S \mathbf{S} \right)
  \nonumber
  \\ 
  &=
  \frac{{\mathcal B}}{2} \left(\gamma_S -\gamma_I \right) ~\,
  \begin{blockarray}{*{2}{c}}
    \begin{block}{*{2}{>{$\footnotesize}c<{$}}}
      $\left\langle \mathrm{S_0}\right|$ &  $\left\langle \mathrm{T_0}\right|$  \\
    \end{block}
    \begin{block}{(*{2}{c})}
      0  & 1   \\
     1 & 0   \\
    \end{block}\\
  \end{blockarray}\label{Zeeman_b}
  ~~,
\end{align}
where $\gamma_I$ and $\gamma_S$ are respective gyromagnetic ratios. The effective Hamiltonian (\ref{rank1_a}) for the parity-nonconserving antisymmetric $J$-coupling in the basis of the two considered states
has the form 
\begin{align}\label{PNC_a}
  H_{\mathrm{PNC}} &=
2\pi
  \frac{i \langle \lambda_z\rangle (\mathcal{E}) }{2}  J^{(1),\mathrm{PNC}} \left(I_+ S_- - I_- S_+ \right)
 \\ \label{PNC_b}
  &=
2\pi
  \frac{\langle \lambda_z\rangle (\mathcal{E}) }{2} J^{(1),\mathrm{PNC}} ~\,
  \begin{blockarray}{*{2}{c}}
    \begin{block}{*{2}{>{$\footnotesize}c<{$}}}
      $\left\langle \mathrm{S_0}\right|$ &  $\left\langle \mathrm{T_0}\right|$  \\
    \end{block}
    \begin{block}{(*{2}{c})}
      0  & i   \\
     -i & 0   \\
    \end{block}\\
  \end{blockarray}
  ~~.
\end{align}
Hence, the total Hamiltonian for the system can be written as
\begin{equation}\label{total_Hamiltonian}
 H =\,\,
  \begin{blockarray}{*{2}{c}}
    \begin{block}{*{2}{>{$\footnotesize}c<{$}}}
      $\left\langle \mathrm{S_0}\right|$ &  $\left\langle \mathrm{T_0}\right|$  \\
    \end{block}
    \begin{block}{(*{2}{c})}
      0  & \frac{\gamma_{\rm eff}}{2}\mathcal{B} + i\frac{D^{\rm PNC}_{\rm eff}}{2}\mathcal{E}  \\
     \frac{\gamma_{\rm eff}}{2}\mathcal{B} - i\frac{D^{\rm PNC}_{\rm eff}}{2}\mathcal{E} & \Delta   \\
    \end{block}\\
  \end{blockarray} \ \ \ .
  \vspace{-12pt} 
\end{equation}
Here we have introduced the effective gyromagnetic ratio $\gamma_{\rm eff} = \gamma_S - \gamma_I$ and the effective PNC-induced E1 coupling $D^{\rm PNC}_{\rm eff} = 2 \pi J^{\rm (1),PNC} \lambda_{z1} /\mathcal{E}_p$.

\section{Measurement scheme}
\subsection{The simplest case, $J=0$}

The basic idea of the proposed measurement scheme is to determine the effective PNC-enabled E1 transition strength, $D^{\rm PNC}_{\rm eff}$, between the states $|\mathrm{S_0}\rangle$ and $|\mathrm{T_0}\rangle$.  This can be accomplished by applying an $\mathcal{E}$-field oscillating at frequency $\omega = \Delta$, which will resonantly drive the $|S_0\rangle \leftrightarrow |T_0\rangle$ transition.  Measuring the Rabi frequency associated with this drive will determine $D^{\rm PNC}_{\rm eff}$ and hence $J^{\rm (1) PNC}$.  As in many other prior or proposed schemes to measure PNC in atoms, the observable effect of the PNC-enabled transition can be amplified through interference with a larger, parity-allowed transition amplitude \cite{Bouchiat_1997}. Here, we use the M1 transition between $|\mathrm{S_0}\rangle$ and $|\mathrm{T_0}\rangle$ for this purpose.  


\begin{figure}[htb]
	\centering
	\includegraphics[width=\columnwidth]{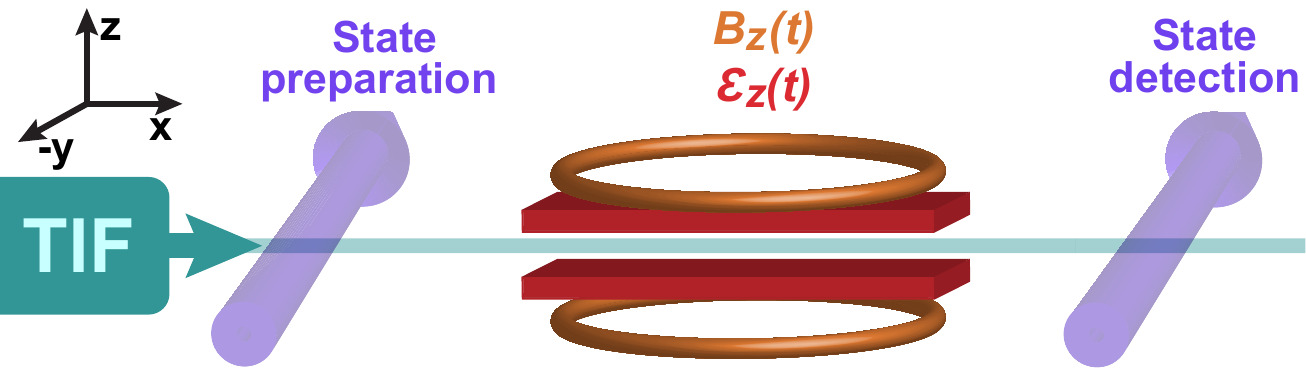}	
	\caption{
	Schematic diagram of the proposed experiment.}
	\label{fig:Schematic} 
\end{figure}

The beam experiment with TlF molecules may look as follows.
We prepare the molecule in the singlet state $|\mathrm{S_0}\rangle$, so that the wavefunction of the two-level system is
$\Psi_0=\bigl(\begin{smallmatrix} 
1\\ 0
\end{smallmatrix}\bigr)$.
Then, each molecule in the
beam
passes through an interaction region, where oscillating electric and magnetic fields are applied along the $z$ axis:
${\cal E}_z (t)={\cal E}_1\cos \omega t$ and ${\cal B}_z (t)={\cal B}_1\sin \omega t$. We write the time-dependent wave function as
$\Psi(t)=\Bigl(\begin{smallmatrix} 
a(t)\\ b(t)
\end{smallmatrix}\Bigr)$.
A given molecule enters the interaction region at time $t_0$ then, at time $t_0+T$, exits; we then measure the population $|b(t_0+T)|^2$ of the triplet state $|\mathrm{T_0}\rangle$.

Here, for simplicity, we describe what happens to a monokinetic slice of molecules. (In reality there will be a distribution of molecular velocities, and one would need to average the results correspondingly.)
Assuming that we are near the resonance, such that the detuning $\delta = \omega - \Delta $ satisfies $|\delta| \ll |\Delta|$, and applying the rotating wave approximation, integrating the Schr\"{o}dinger equation with the Hamiltonian
\eqref{total_Hamiltonian} from $t_0$ to $t_0+T$ yields the result for the population of the triplet: 
\begin{align}
    \label{population}
    &|b(t_0+T)|^2 = \frac{\Omega_{\rm M1}^2}{2\Omega^{\prime^2}}\left(1-\cos{\Omega^\prime T}\right)
    \\ \nonumber
    &-\frac{\Omega_{\rm PNC}\Omega_{\rm M1}}{\Omega^{\prime^2}} \left[\frac{\Omega_{\rm M1}^2T}{2\Omega^\prime}\sin{\Omega^\prime T}
    +\frac{\delta^2}{\Omega^{\prime^2}}\left(1-\cos{\Omega^\prime T}\right)\right],
\end{align}
independent of $t_0$.
Here,
\begin{align}
    \label{PNC}
    \Omega_{\rm PNC} &=\frac{D^{\rm PNC}_{\rm eff}}{2}\mathcal{E}_1,
    \\ \label{Gamma}
    \Omega_{\rm M1} &= \frac{\gamma_{\rm eff}}{2}{\cal B}_1,
    \\ \label{Omega}
    \Omega^\prime &= \sqrt{\delta^2+\Omega_{\rm M1}^2},
\end{align}
and we have discarded the small terms quadratic in $\Omega_{\rm PNC}$.
The first term in Eq.\,\eqref{population} is independent of $\Omega_{\rm PNC}$, and the second one is linear in $\Omega_{\rm PNC}$. At resonance, where $\delta=0$ and $\Omega^\prime = \Omega_{\rm M1}$, this expression simplifies to
\begin{align}
    \label{population_res}
    |b(t_0 + T)|^2 &= \frac12 \left(1-\cos{\Omega_{\rm M1} T}\right)
    -\frac{\Omega_{\rm PNC}T}{2}\sin{\Omega_{\rm M1} T}\,.
\end{align}

Throughout this discussion, we have assumed that the level splitting $\Delta$ remains constant. However, Stark-induced mixing of the $J=0$ state with other rotational levels, in combination with hyperfine couplings, causes the value of $\Delta$ to change as a function of $\mathcal{E}$: approximately, $\Delta$ in the expressions above should be replaced by
\begin{align}
    \Delta(J=0) = \Delta_0 + \alpha\mathcal{E}^2 = \Delta_0 + \alpha \mathcal{E}_1^2 \cos^2(\omega t) \\
    = \Delta_0 + \frac{\alpha}{2}\mathcal{E}_1^2 + \frac{\alpha}{2}\mathcal{E}_1^2 \cos(2\omega t)\, .
\end{align}
Under realistic conditions (as discussed below), the frequency modulation depth $\alpha \mathcal{E}_1^2/2$ can be large compared to the linewidth due to time-of-flight, $\delta\omega \approx 1/T$. (The DC offset $\alpha/2$ can be accounted for simply by a small change of the driving frequency $\omega$.) It is not immediately obvious that this deep modulation does not dramatically change the outcome described above, when zero modulation was assumed.
\begin{figure}[htb]
    \centering
    \includegraphics[width=\columnwidth]{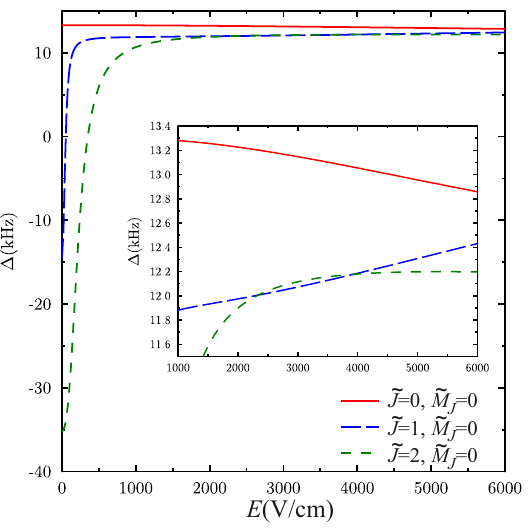}
    \caption{Splitting $\Delta$ for three lowest rotational levels with $\tilde{M}_J=0$. 
}
    \label{fig:Delta}
\end{figure}

In fact, under the conditions we consider here and below, this rapid but seemingly large oscillation of $\Delta$ has negligible effect on the transition probability.  This can be seen as follows. Modulation of the resonant frequency $\Delta$ is equivalent to an equal-amplitude (but opposite sign) modulation of the driving frequency $\omega$.  A sinusoidal drive frequency modulation of the form $\omega(t) = \omega_0 + \kappa \sin{\omega_{\rm mod} t}$ is equivalent to a phase modulation, such that the frequency-modulated oscillation $\sin{[\omega(t) t]}$ can be replaced by $\sin{[\omega_0 t + (\kappa/\omega_{\rm mod}) \cos{\omega_{\rm mod} t }] }$.  In our case, the phase modulation frequency is $\omega_{\rm mod} = 2\omega_0$, and the phase modulation index is $\beta = \alpha \mathcal{E}_1^2/(4 \omega_0)$.  This is equivalent to driving with a comb of frequencies $\omega_n = \omega_0 \pm n\omega_{\rm mod} = \omega_0(1 \pm 2n)$, with corresponding amplitudes $A_n \propto J_n(\beta)$, where $J_n$ is the Bessel function of order $n$ (see e.g.~\cite{budker2008atomic}).  When $\beta \ll 1$, i.e. when the frequency modulation depth satisfies $\alpha \mathcal{E}_1^2/4 \ll \omega_0$, the amplitude of the fundamental frequency is $J_0(\beta) = 1 + \mathcal{O}(\beta^2)$, and that of the $n^{\rm th}$ sideband is $J_n(\beta) \sim \beta^n \ll 1$. Hence the modulation has negligible effect on the amplitude of the resonant component of the driving field, and only introduces small-amplitude sidebands of far off-resonant driving. That is, its effect is negligible in our scheme.

Now, we return to the main discussion.  The P-odd signal arising from Eq.~(\ref{population_res}) corresponds to the change in the population of the triplet and singlet levels when we reverse the sign of either one (but not both) of the fields $\mathcal{E}_1$ or $\mathcal{B}_1$, since $\Omega_{\rm PNC} \propto {\cal E}_1$ and $\Omega_{\rm M1} \propto {\cal B}_1$. Fluctuations of the population (the first term independent of $\Omega_{\rm PNC}$) give the noise. Both the signal and the noise turn to zero for $\Omega_{\rm M1} T= 2\pi n$; the signal is maximal when $\Omega_{\rm M1} T= \pi(n+\tfrac12)$.  We consider the case where both the triplet state population $P_T = |b(t_0+T)|^2$ and the singlet state population $P_S = 1 - P_T$ are detected, analogous to the plans for detecting both spin quadratures after the Ramsey spin rotation protocol in CeNTREX~\cite{CeNTREX21}. In this case, the PNC signal-to-noise ratio is:  
\begin{align}\label{S/N}
   \mathrm{S/N}=
   \Omega_{\rm PNC} T\sqrt{N}\,,    
\end{align}
where $N$ is the number of detected molecules.

This type of experiment is similar to what was discussed for atomic hydrogen in the 1970s \cite{Lewis1975,Dunford1978} and later for alkali metals \cite{GEKM88,Gomez2007FrAnapoleScheme}.  In these papers, the idea was to look for interference between a PNC-induced E1 amplitude and a parity-allowed M1 amplitude, driven on resonance between hyperfine sublevels of the same state. Because time-reversal (T) symmetry is conserved, and hence the ratio of the M1 and PNC E1 amplitudes is pure imaginary \cite{bouchiat1974parity}, these amplitudes can only interfere when the driving $\mathcal{E}$- and $\mathcal{B}$-fields are $\pi/2$ out of phase, as here. Our scheme is also closely related to the proposal by Fortson to measure E1$_{\rm PNC}$-E2 interference in a single trapped ion \cite{Fortson93_Ion_PNC}. As discussed there, the S/N is at the standard quantum limit for measuring the AC Stark shift $\Omega_{\rm PNC}$, for $N$ particles observed over coherence time $T$. Because the resonant frequency $\Delta$ is so small here (as compared to all cases in the prior literature), both applied fields are entirely in the near-field regime, with control afforded by simple patterns of conductors and ordinary function generators. We also note that despite some superficial resemblances, the proposed method here is quite different from those used to measure PNC effects in near-degenerate states in Dy \cite{Nguyen1997} and in diatomic molecules \cite{DeMille:2008} such as BaF \cite{AACD18}. In those experiments, the pair of levels considered have opposite parity rather than the same parity as here; the level splitting $\Delta$ can be tuned through zero, whereas here it is fixed; and the field driving the system is at a frequency far above resonance, rather than on resonance as here.

\subsection{The $J>0$ case}
The measurement scheme as just described is applicable only to the spin sublevels of the $J=0$ rotational state.  However, it is highly desirable to also perform analogous measurements in states with larger values of $J$.  The reason is that in such states, the slope $\lambda_{z1}$ of the plot of polarization $\langle \lambda_z \rangle$ vs.\,$\mathcal{E}$ (see Fig.\,\ref{fig:n_z}) can be of opposite sign to that in the $J=0$ state.  This can provide a powerful means to detect and minimize important systematic errors (see Sec.\,\ref{Sec:Considerations_TlF}).
However, for $J>0$ states in zero $\mathcal{E}$-field, spin-rotation couplings arising from the terms proportional to $c_1,c_2,$ and $c_3$ in Eq.\,\eqref{H_HFS} result in energy eigenstates that are not well-described as spin singlet and triplet states. In this regime, our present treatment does not apply and its consideration goes beyond the scope of the present work. However, the triplet-singlet state character can be recovered in the presence of a substantial $\mathcal{E}$-field. Hence, to perform an analogous PNC measurement in states with $J>0$, we must consider the situation where not only the near-resonant oscillating field $\mathcal{E}_1 \cos{\omega t}$, but also a large static field $\mathcal{E}_0$, is present. We consider this situation here. Though most of the discussion will be valid for any states with $J>0$, for concreteness we focus on the case of $J=1$. 

In the presence of a sufficiently large polarizing $\cal{E}$-field, such that the Stark energy $d\mathcal{E}$ is large compared to the spin-rotation couplings (arising from the terms proportional to $c_1,c_2,$ and $c_3$ in Eq.~(\ref{H_HFS})), the spins $S$ and $I$ can decouple from the rotation. In this regime, the vector $\bm \lambda$ rapidly precesses around the electric field, $\bm{\mathcal{E}}$, and the $\mathcal{E}$-field mixes rotational levels of opposite parity (corresponding to $J$ even or odd).

The Stark-split energy eigenstates in the large field regime can be described in terms of approximate quantum numbers $\tilde{J},\tilde{M}_J$. Here, $\tilde{J}$ corresponds to the value of $J$ that any given state connects to, if the $\mathcal{E}$-field were adiabatically reduced to zero \cite{CeNTREX21}; $\tilde{M}_J$ corresponds to the value of $M_J$ in the limit where all spin-rotation constants vanish. In this regime, states with $\tilde{M}_J=0$, for any value of $\tilde{J}$, have no diagonal matrix elements of the spin-rotation interactions (to first order in the small ratio $c_j / d\mathcal{E}$). Instead, here---to first order---hyperfine structure is determined by the spin-spin interaction [i.e. the term proportional to $c_4$ in Eq.~(\ref{H_HFS})], just as in the field-free $J=0$ state described above.  Hence here, as before, the spin configurations are (to good approximation) singlet and triplet states.

To remain in the large-field regime, the magnitude of the $\cal{E}$-field should always remain sufficiently large.  To ensure this in the presence of the oscillating $\cal{E}$-field used to drive the PNC-induced E1 singlet-triplet transition, it is necessary to simultaneously apply a larger DC field. That is: we consider the case where the total applied $\cal{E}$-field has the form $\mathcal{E}_0 + \mathcal{E}_1\cos{\omega t}$, where $\mathcal{E}_0 > |\mathcal{E}_1|$. In practice, for the $\tilde{J}=1,\tilde{M}_J=0$ states in TlF, it is sufficient to maintain $\mathcal{E}_0 - |\mathcal{E}_1| \gtrsim 500$ V/cm.

The primary complication to the PNC measurement arising from the static field $\mathcal{E}_0$ is its effect on the singlet-triplet level splitting $\Delta$. As mentioned earlier, the presence of $J\neq 0$ components of the wavefunction of every eigenstate---in combination with off-diagonal couplings, due to spin-rotation interactions, with states where $\tilde{M}_J \neq 0$---leads to a shift in $\Delta$.  Because the admixtures of different $J$ states into a given $\tilde{J},\tilde{M}_J=0$ state change with the size of the applied $\mathcal{E}$-field, $\Delta$ also changes with $\mathcal{E}$. The dependence of $\Delta$ on $\mathcal{E}$, determined by numerical diagonalization of the full Hamiltonian, is shown in Fig.~\ref{fig:Delta} for states with $\tilde{M}_J=0$ and $\tilde{J}=0,1,2$. This illustrates clearly the very strong dependence of $\Delta$ on $\mathcal{E}$ when $\mathcal{E}$ is small and $\tilde{J}>0$. However, the dependence weakens for small ranges around sufficiently large values of $\mathcal{E}$. Here, we consider the regime where, in the range $\mathcal{E}_0 - |\mathcal{E}_1| < \mathcal{E} < \mathcal{E}_0 + |\mathcal{E}_1|$, $\Delta(\mathcal{E})$ is approximately a linear function:
\begin{align}
    \Delta(\mathcal{E}_0 + \mathcal{E}_1) \approx \Delta_{\mathcal{E}0} + \alpha^\prime \mathcal{E}_1\,,
\end{align}
where $\Delta_{\mathcal{E}0} \equiv \Delta(\mathcal{E}_0)$.
From Fig.\,\ref{fig:Delta}, for the states with $\tilde{M}_J=0$ and $\tilde{J}=1$ ($\tilde{J}=0$) in TlF, this is a good approximation for
$\mathcal{E}_0 \approx 3000$\,V/cm 
and $|\mathcal{E}_1| = 1000$\,V/cm.  
Here, $\Delta_{\mathcal{E}0} \approx 12.1$\,kHz (13.1\,kHz) and $\alpha^\prime \approx 0.1$\,kHz/(kV/cm) ($-$0.1\,kHz/(kV/cm))
for the $\tilde{J} = 1$ $(\tilde{J} = 0)$ state. Note also that in this range, the molecular polarization $\langle \lambda_z \rangle$ is approximately linear in $\mathcal{E}$ over the entire range (see Fig.~\ref{fig:n_z}).

The Hamiltonian $\tilde{H}$ for this system can be written, in analogy to Eq.~(\ref{total_Hamiltonian}), as
\begin{widetext}
\begin{equation}\label{tilde_Hamiltonian}
 \tilde{H} =\,\,
  \begin{blockarray}{*{2}{c}}
    \begin{block}{*{2}{>{$\footnotesize}c<{$}}}
      $\left\langle \mathrm{\tilde{S}_0}\right|$ &  $\left\langle \mathrm{\tilde{T}_0}\right|$  \\
    \end{block}
    \begin{block}{(*{2}{c})}
      0  & \frac{\gamma_{\rm eff}}{2}\mathcal{B}_1\sin{\omega t} + i\frac{D^{\rm PNC}_{\rm eff}}{2}[\mathcal{E}_0 +\mathcal{E}_1\cos{\omega t}]  \\
     \frac{\gamma_{\rm eff}}{2}\mathcal{B}_1\sin{\omega t} - i\frac{D^{\rm PNC}_{\rm eff}}{2}[\mathcal{E}_0 +\mathcal{E}_1\cos{\omega t}]  & \Delta_{\mathcal{E}0} +\alpha^\prime \mathcal{E}_1\cos{\omega t}   \\
    \end{block}\\
  \end{blockarray} \ \ \ ,
\end{equation}
\end{widetext}
where the eigenstates $\left\langle \mathrm{\tilde{S}_0}\right|$ and $\left\langle \mathrm{\tilde{T}_0}\right|$ here are the singlet and triplet states when $\mathcal{E} = \mathcal{E}_0$ and the PNC effect is absent. Note that, in this more general situation, the factor 
$\lambda_{z1}/\mathcal{E}_p$
that appears in the definition of $D^{\rm PNC}_{\rm eff}$ is replaced by the slope of $\langle \lambda_z \rangle$ vs.~the applied $\mathcal{E}$-field, at the bias field $\mathcal{E}_0$.  
That is, now $\lambda_{z1}/\mathcal{E}_p \rightarrow d\langle \lambda_z \rangle / d\mathcal{E}$, evaluated at $\mathcal{E}=\mathcal{E}_0$. As mentioned above, this slope can be of opposite sign in the $\tilde{J}=1$ state and the $\tilde{J}=0$ state, so that $D^{\rm PNC}_{\rm eff}$ is of opposite sign in these states.

Importantly, the off-diagonal matrix elements related to the $\mathcal{E}$-field include \textit{only} PNC couplings. This fact seems far from obvious, given that the eigenstates here are mixed-parity states. A proof of the remarkable fact that ordinary $\mathcal{E}$-field induced mixing of these states vanishes is given in Appendix~\ref{AppNoE1}.

Another important observation is that the occupation of the triplet (or singlet) state which is the module square of the solution of the Schr\"{o}dinger equation with the Hamiltonian~(\ref{tilde_Hamiltonian}) is an even function of $\mathcal{B}_1$ ($\Omega_{M1}$) in the absence of PNC effects. This directly follows from the form of the equation. Thus, the modulation of $\Delta_{\mathcal{E}0}$ should not lead to systematic effect that imitates PNC signal which is an odd function of $\mathcal{B}_1$ ($\Omega_{M1}$).

As before, by driving the system on resonance $(\omega = \Delta_{\mathcal{E}0})$, applying the rotating wave approximation, and writing $\tilde{H}$ in the rotating frame, this simplifies to
\begin{equation}\label{total_Hamiltonian3}
 \tilde{H}_{\rm rot} =\,\,
  \begin{blockarray}{*{2}{c}}
    \begin{block}{*{2}{>{$\footnotesize}c<{$}}}
      $\left\langle \mathrm{\tilde{S}_0}\right|$ &  $\left\langle \mathrm{\tilde{T}_0}\right|$  \\
    \end{block}
    \begin{block}{(*{2}{c})}
      0  & -\frac{i}{2} \Omega_{\rm M1} + \frac{i}{2}\Omega_{\rm PNC}  \\
     \frac{i}{2}\Omega_{\rm M1} - \frac{i}{2}\Omega_{\rm PNC}  & \alpha^\prime \mathcal{E}_1\cos{\omega t}   \\
    \end{block}\\
  \end{blockarray} \ \ \ .
  \vspace{-12pt} 
\end{equation}
The sinusoidal modulation of the level splitting, as before, is equivalent to driving the system with small sidebands at frequencies (here at $\omega_n = \omega \pm n\omega$), and leads to negligible effects. Hence, finally, this system in the presence of the static field $\mathcal{E}_0$ and for a state with any value of $\tilde{J}$ is entirely equivalent to the simpler version discussed before, and the PNC signal can be measured in the same way.


\section{\textit{Ab initio} calculation of $J$-coupling parameters in TlF}\label{Sec:abinitio}
\subsection{Electronic Structure Theory}\label{Sec:EST}

Let us now consider parameters that are required to estimate the expected PNC signal (\ref{S/N}). Different sources of interactions between the nucleus and electrons in the molecule 
can induce $J$-coupling (\ref{tensor}). Here we are interested in two of them --- the P-even interaction of the electrons with the nuclear magnetic dipole moment and the P-odd interaction with the anapole moment of the nucleus \cite{Zel57t}. The former one has the form:
 \begin{align}\label{HF-rel}
 {H}_\mathrm{HF, K}
 &= \sum_{p}
 \gamma_K \frac{\mathbf{I}_K \cdot\ [(\mathbf r_p -\mathbf R_K) \times \bm \alpha_p]}
 {|\mathbf r_p -\mathbf R_K|^3}\,,
 \end{align}
where summation goes over all electrons of the molecule, $\mathbf R_K$ is the position of the nucleus, $\mathbf r_p$ is the position of electron $p$, $\bm \alpha_p$ are Dirac matrices. The P-odd interactions inside the nucleus $K$ can induce the anapole moment that can be characterized by dimensionless constant $g_K$. The PNC interaction between the nucleus and electrons induced by this moment is~\cite{GFreview}:
 \begin{align}
 \label{NSD-PNC-rel}
 {H}_\mathrm{P,K}
 &=\frac{G_F}{\sqrt2}
 \sum_{p} g_K \bm \alpha_p\mathbf{I}_K
 \rho_K(\mathbf r_p-\mathbf R_K)\,,
 \end{align}
where $G_\mathrm{F}~=~2.22249 \cdot 10^{-14}$~a.u. is the Fermi coupling constant in atomic units and $\rho_K(\bm r)$ is the charge density of the nucleus. In addition to the anapole moment there are other contributions to the interaction \eqref{NSD-PNC-rel}. These contributions can be accounted for by redefining the coupling constant $g_K$ \cite{GFreview}.

Interactions (\ref{HF-rel}) and (\ref{NSD-PNC-rel}) can contribute to the effective Hamiltonian~(\ref{tensor}) in different combinations. One of them is the parity conserving indirect coupling of magnetic moments of the nuclei $A$ and $B$ through electronic shells via interactions~(\ref{HF-rel});
it will be designated as $J^{\mathrm{NMR}}$. Its components can be calculated as the mixed derivative of the molecular energy $E$ with respect to $I$ and $S$ and with $g_I=0$, $g_S=0$:
\begin{equation}
\label{JcouplingNMR}
\left.2\pi J^{\mathrm{NMR}}_{ik}=\frac{\partial^2E}{\partial I_{i}\partial S_{k}} \right|_{g_I=0, g_S=0}.
\end{equation}
In the case of a diatomic molecule oriented along axis $z$, the tensor $J^{\mathrm{NMR}}_{ik}$ has two unique components, $J^{\mathrm{NMR}}_{\perp}=J^{\mathrm{NMR}}_{\rm xx}=J^{\mathrm{NMR}}_{\rm yy}$ and $J^{\mathrm{NMR}}_{||}=J^{\mathrm{NMR}}_{\rm zz}$. It is convenient to use their isotropic combination:
\begin{equation}
\label{Jiso}
J^{\mathrm{NMR}}_{iso}=(J^{\mathrm{NMR}}_{\rm ||}+2J^{\mathrm{NMR}}_{\perp})/3
\end{equation}
and anisotropy: 
\begin{equation}
\label{Janiso}
\Delta J^{\mathrm{NMR}}=J^{\mathrm{NMR}}_{\rm ||}-J^{\mathrm{NMR}}_{\perp}.
\end{equation}
$J^{\mathrm{NMR}}_{iso}$ is the scalar (rank-0) $J$-coupling part of (\ref{tensor});
note that $J^{\mathrm{NMR}}_{iso}=c_4$ in Eq.\,(\ref{H_HFS}). The $\Delta J^{\mathrm{NMR}}$ constant can be obtained from the experimental value of $c_3$ constant~\cite{Wilkening:1984,Boeckh1964} by subtracting the contribution of the direct interaction of the magnetic dipole moments of Tl and F nuclei, which has been done in Ref.~\cite{Bryce:2000}.

Components of the $J^{\mathrm{PNC}}_{ik}$ tensor that characterise the PNC contribution induced by the anapole moment of the nucleus $A$ to $J_{ik}$ in Eq.~(\ref{tensor}) can be calculated as:
\begin{equation}
\label{JcouplingPNC}
2\pi \left.J^{\mathrm{PNC}}_{ik}=\frac{\partial^2E}{\partial I_{i}\partial S_{k}} \right|_{\gamma_{I}=0,g_{S}=0}.
\end{equation} 
In TlF  $\mathbf{J}^{(1),\mathrm{PNC}}$ is collinear with the molecular axis. The $J^{\mathrm{PNC}}_{ik}$ tensor is antisymmetric. It follows from Eqs.\ \eqref{tensor}-\eqref{rank1} that $J^{(1),\mathrm{PNC}}_z=J^{\mathrm{PNC}}_{xy}$.

\subsection{Computational details}

In the one-particle case and Dirac theory, $J^{\mathrm{NMR}}_{ik}$ can be calculated using the operator \eqref{HF-rel} within the sum-over-states approach:
 \begin{eqnarray}
 \label{Jnmrij}
& & 2\pi J^{\mathrm{NMR}}_{ik}\\ 
& & =\sum_n
 \frac{\langle 0 | \gamma_I \frac{[(\mathbf r -\mathbf R_I)\times \bm \alpha]_i} {|\mathbf r -\mathbf R_I|^3}  |n \rangle
       \langle n|   \gamma_S   \frac{[(\mathbf r -\mathbf R_S) \times \bm\alpha]_k} {|\mathbf r -\mathbf R_S|^3}     |0 \rangle}
      {E_0-E_n} + c.c. \nonumber
 \end{eqnarray}
In this equation, the sum goes over positive and negative energy states, excluding the occupied one-particle state $|0 \rangle$. Below we will distinguish the positive and negative energy parts of Eq.~(\ref{Jnmrij}). $J^{\mathrm{PNC}}_{xy}$ can be calculated in a similar way for the one-particle case:
 \begin{multline}
 2\pi J^{\mathrm{PNC}}_{xy} \\ 
 =\sum_n
 \frac{\langle 0 | \frac{G_F}{\sqrt2}
 g_I \alpha_x
 \rho_I(\mathbf r-\mathbf R_I) |n \rangle
       \langle n| \gamma_S  \frac{[(\mathbf r -\mathbf R_S) \times \bm \alpha]_y}
 {|\mathbf r -\mathbf R_S|^3}     |0 \rangle}
      {E_0-E_n} + c.c. 
 \,
 \label{Jxy}
 \end{multline}

The simplest generalization of Eqs.~(\ref{Jnmrij}) and (\ref{Jxy}) on the many-electron case requires to sum also over all occupied molecular bispinors, i.e. replace $|0\rangle$ by the sum over all occupied $|i\rangle$. This is the co-called uncoupled approximation and will be referenced as PT2. In a more accurate treatment, one can use the linear response technique developed for both Dirac-Hartree-Fock (DHF) and density functional theory (DFT) approaches to calculate components of $J$-coupling and other NMR properties~\cite{Weijo:2007,Bast:2006,Olejniczak:12,Ilias:13,Aucar:99,DIRAC15}. This approach corresponds to the ``analytical'' treatment of derivatives (\ref{JcouplingNMR}) and (\ref{JcouplingPNC}) at the DHF and DFT levels instead of simple summations in (\ref{Jnmrij}) and (\ref{Jxy}).
%
%
For example, within the linear-response DHF method~\cite{Aucar:99}, one obtains the following closed-form expression for $J^{\mathrm{PNC}}_{xy}$: 
\begin{widetext}
 \begin{multline}
\begin{aligned}  
 2\pi J^{\mathrm{PNC}}_{xy} =\sum_{k,l,a,b}
 \langle k | \frac{G_F}{\sqrt2}
 g_I \alpha_x
 \rho_I(\mathbf r-\mathbf R_I) |a \rangle P^{-1}_{a,k;b,l}
       \langle b| \gamma_S  \frac{[(\mathbf r -\mathbf R_S) \times \bm \alpha]_y}
 {|\mathbf r -\mathbf R_S|^3}     |l \rangle
       + c.c. 
      \nonumber
 \, ,
\end{aligned} 
 \end{multline}
 \end{widetext}
where
 \begin{multline}
 \nonumber
 \mathbf{P^{-1}}=\begin{bmatrix}
\mathbf{A} & \mathbf{B^*}\\ 
\mathbf{B} & \mathbf{A^*} 
\end{bmatrix}^{-1}, \\
A_{k,l,a,b}=\delta_{a,b}\delta_{k,l}(E_a - E_k)+\langle al|g|kb \rangle-\langle al|g|bk \rangle, \\
B_{k,l,a,b}=\langle kl|g|ab\rangle-\langle kl|g|ba\rangle.
 \end{multline}
In the expression above, $k,l$ run over occupied orbitals, $a,b$ run over virtual positive and negative-energy orbitals, $E_k, E_a$ are corresponding orbital energies, $g$ is the electron-electron Coulomb operator.
The positive-energy states contribution can be obtained by restricting $a,b$ above by the positive-energy virtual orbitals only. We define and calculate the ``negative-energy contribution'' in the many-electron case as the difference between the total linear-response value (i.e. with no restrictions applied on indexes $a,b$) for derivatives (\ref{JcouplingNMR}) or (\ref{JcouplingPNC}) and the corresponding positive-energy states contribution. One can see that neglecting two-electron integrals $\langle \dots|g|\dots \rangle$ in the expression above leads to the PT2 approach.
Below we will call 
the linear-response DHF or DFT methods
just DHF or DFT. The positive energy part of the related property, a shielding tensor for a molecule containing a heavy atom, can be calculated with a smaller uncertainty than DHF/DFT within the relativistic coupled cluster theory~\cite{Skripnikov:18a,Skripnikov:2020a,Skripnikov:2022a}. Here we have generalized this approach to calculate PNC and P-conserving contributions to the nuclear indirect spin-spin coupling. For this we have numerically calculated mixed derivatives (\ref{JcouplingNMR}) and (\ref{JcouplingPNC}) for the energy $E$ obtained within the coupled cluster theory. In the used procedure we obtained a set of occupied and virtual orbitals within the molecular Dirac-Hartree-Fock method for TlF. After that, the interactions (\ref{HF-rel}) and (\ref{NSD-PNC-rel}) were added to the electronic Hamiltonian and coupled cluster calculations were performed to obtain the positive energy contribution. 

All molecular calculations have been performed within the Dirac-Coulomb Hamiltonian. We have used the combinations of Dyall's Gaussian-type basis sets. For example, the TZTZ basis set corresponds to the uncontracted Dyall's AAETZ  basis set for both Tl and F atoms \cite{Dyall:07,Dyall:12,Dyall:2016}, while TZQZ corresponds to the AAETZ basis set for Tl and AAEQZ for F, etc. In all-electron (90e) correlation calculations within the coupled cluster with single and double cluster amplitudes method (CCSD), we have not used any truncation of the virtual orbitals by their energies. To consider the contribution of high-order correlation effects at the level of coupled cluster with single, double, and triple amplitudes (CCSDT) we have performed correlation calculations for 20 valence electrons and set the cutoff for virtual orbitals energies to 30 $E_H$, and used the DZDZ basis with the extended number of s- and p- type functions for both Tl and F. This basis set will be called DZDZext. For calculation of the negative energy contribution, we have also used the relativistic density functional theory with the hybrid Perdew-Burke-Ernzerhof PBE0 functional~\cite{pbe0}. 

For the problem under consideration, the basis set for all-electron coupled cluster calculations should include functions with large exponential parameters to accurately reproduce the wavefunction asymptotic near and inside the nucleus. The most important here are s$_{1/2}$- and p$_{1/2}$-type functions which can penetrate inside the nucleus. We have found that for the fluorine atom the saturation is achieved if one includes $s$ and $p$ basis functions corresponding to the AAEQZ~\cite{Dyall:07,Dyall:12,Dyall:2016} basis set. Therefore, we have used this basis set for fluorine in the main calculations of $J$-coupling contributions. At the same time for the thallium atom rather accurate description is achieved already for the AAETZ~\cite{Dyall:07,Dyall:12,Dyall:2016} basis set.

In all calculations, the experimental internuclear distance R(Tl--F)=3.94 Bohr~\cite{Huber:1979} has been employed. For the nuclear magnetic moments we used the values from Ref.~\cite{Sto05}: $\mu(^{205}{\rm Tl}) = 1.63821~\mu_N$, $\mu(^{19}{\rm F}) = 2.62887~\mu_N$.

We have used the Gaussian model for the nuclear charge distribution which is well suited for molecular calculations~\cite{Visscher:1997}. Relativistic four-component calculations have been performed within the locally modified {\sc dirac} \cite{DIRAC15,Saue:2020} and {\sc mrcc} \cite{MRCC2020} codes. For calculation of matrix elements of operators (\ref{HF-rel}) and (\ref{NSD-PNC-rel}) the code developed in  Refs.~\cite{Skripnikov:16b,Skripnikov:15b,Skripnikov:15a} was used.

\subsection{Results and discussion of calculation}
Table\,\ref{TconservingJ} gives the calculated values of the parity conserving indirect nuclear spin-spin coupling tensor components (\ref{Jiso}) and (\ref{Janiso}) in comparison with the experimental values~\cite{Wilkening:1984,Bryce:2000,Boeckh1964}. We have found that the negative energy states contributions to $J^{\mathrm{NMR}}_{ik}$ calculated within the linear response DHF, DFT or uncoupled DHF (PT2) coincide within a few Hz. Similar behavior has been found for the shielding constant~\cite{Skripnikov:18a}. However, the positive energy states contribution strongly depends on the level of the theory to treat correlation effects. It can be seen that the value of the anisotropy obtained within the density functional theory deviates from the experimental value by about 20\%. Therefore, we have used the relativistic 4-component coupled cluster approach to calculate the positive energy P-conserving contribution. As it can be seen from Table\,\ref{TconservingJ} such an approach provides rather accurate values of both isotropic and anisotropic components of $J^{\mathrm{NMR}}$.

\begin{table}[htb]
\caption{Isotropic and anisotropic components (in Hz) of P-conserving indirect nuclear spin-spin coupling in ${\rm ^{205}Tl^{19}F}$.}
\begin{tabular}{lrr}
\hline
\hline
   Contribution            & $J^{\rm NMR}_{iso}$ & ~$\Delta J^{\rm NMR}$\\
\hline                 
\multicolumn{3}{c}{Negative energy states contributions}            \\ 
DHF-uncoupled(PT2)/QZQZ                     & 0               & 189         \\
DHF/QZQZ                     & 0               & 188         \\
DFT/QZQZ (total negative)    & 0               & 191         \\
\\
\multicolumn{3}{c}{Positive energy states contributions}            \\
DHF/TZQZ                   &  -27582         &   11647        \\
DFT/TZQZ                   & -15462          &   13123        \\
\\
90e-CCSD/TZQZ              & -13586          & 9820           \\
+20e-(CCSDT-CCSD)/DZDZext      &  122            &  867           \\
Total positive             &  -13463         &  10687         \\
                           &                 &                \\
Total                      & -13463           & 10877          \\
Experiment~\cite{Wilkening:1984,Bryce:2000,Boeckh1964}       & -13300(700)     & 11100(500)   \\
\hline
\hline
\end{tabular}
\label{TconservingJ}
\end{table}


Table~\ref{TpncJ} gives the calculated values of the PNC contribution to the indirect nuclear spin-spin coupling $J^{(1),\mathrm{PNC}}$~ (\ref{Jpnc_a}).

\begin{table}[htb]
\caption{PNC contribution to $J$-coupling, $J^{(1),\mathrm{PNC}}$, in ${\rm ^{205}Tl^{19}F}$ induced by the anapole moment of the $^{205}$Tl nucleus.}
\begin{tabular}{lr}
\hline
\hline
Contribution & Value, $10^{-3} g_{\rm Tl}$ Hz.       \\
\hline
\multicolumn{2}{c}{Negative energy states contributions}       \\
DHF-uncoupled(PT2)/QZQZ           & -2.82  \\
DHF/QZQZ                         & -2.82  \\
DFT/QZQZ  (total negative)       & -2.83  \\
                                 &       \\
\multicolumn{2}{c}{Positive energy states contributions}     \\
DHF/TZQZ                          & -2.05 \\        
DFT/TZQZ                          & -0.52 \\
\\
90e-CCSD/TZQZ                     & -0.13 \\
+ 20e-(CCSDT-CCSD)/DZDZext          & -0.07 \\
Total positive                    & -0.19 \\
                                  &      \\
Total                             & -3.03 \\
\hline
\hline
\end{tabular}
\label{TpncJ}
\end{table}

It can be seen that the leading contribution to this coupling is due to negative energy states. This contribution is practically the same for the linear response DHF, DFT, or uncoupled DHF (PT2). In contrast, the positive energy contribution to $J^{(1),\mathrm{PNC}}$ strongly depends on the level of the electronic correlation treatment, even more strongly than the P-conserving term, e.g. the DFT value is 4 times smaller than the DHF value. The explicit treatment of electron correlation effects within the relativistic coupled cluster approach with single and double amplitudes leads to even stronger suppression of the positive energy contribution. The inclusion of triple cluster amplitudes gives a small, but non-negligible contribution compared with the total positive energy one. It can be seen that the density functional theory considerably overestimates the positive energy contribution. Similar overestimation has been also outlined in Ref.~\cite{Weijo:2007}. Note that the total positive energy contribution calculated at the CC level itself is more than an order of magnitude smaller than the negative energy one. According to our estimates, the final uncertainty of $J^{(1),\mathrm{PNC}}$ given in Table~\ref{TpncJ} is less than 8\%.

The final value of $|J^{(1),\mathrm{PNC}}|$ is of the same order as the estimation obtained in Ref.~\cite{Blanchard:2020}, $9\times10^{-3} g_{\rm Tl}$~Hz. The equation that has been used in Ref.~\cite{Blanchard:2020} to estimate $J^{(1),\mathrm{PNC}}$ can be obtained from the equation similar to Eq.~(\ref{Jxy}) with consideration of only the negative energy contribution and some further approximations~\cite{Blanchard:2020}: setting $|1s({\rm Tl})\rangle$ as $|0\rangle$, replacement of $(E_0 - E_n)$ by $2mc^2$ and replacement of the $\sum_n |n\rangle \langle n|$ by $1$. Similar approach for the diamagnetic contribution to the shielding constant is known as the Sternheim's approximation~\cite{Sternheim:1962,Aucar:99}. 
It is instructive to calculate the following sum:
\begin{eqnarray}
&  & 2\pi J^{\mathrm{PNC}}_{xy}     \\
&  & =\sideset{}{'}\sum_n
  \frac{\langle 1s | \frac{G_F}{\sqrt2}
  g_I \alpha_x
  \rho_I(\mathbf r-\mathbf R_I) |n \rangle
       \langle n| \gamma_S  \frac{[(\mathbf r -\mathbf R_S)\times \bm \alpha ]_y}
  {|\mathbf r -\mathbf R_S|^3}     |1s \rangle}
       {2mc^2} + c.c.  \nonumber 
\end{eqnarray}
where $|1s\rangle$ is the lowest positive-energy molecular bispinor obtained within the Dirac-Hartree-Fock approach and summation over $|n\rangle$ includes only the negative energy states. The obtained value in this approximation, $-8.9\times10^{-3} g_{\rm Tl}$ Hz, almost coincides with the result of Ref.~\cite{Blanchard:2020}. If one further includes summation over all occupied bispinors $|i\rangle$ (instead of only $|1s\rangle$) then the value is $-10.9\times10^{-3} g_{\rm Tl}$~Hz. However, if one further uses the actual values of $(E_i - E_n)$ instead of the $(E_i - E_n)\approx2mc^2$ approximation then we come to the ``PT2'' value for the negative energy contribution given in the first line of Table~\ref{TpncJ}, i.e. about the 3 times smaller value $-2.82\times10^{-3} g_{\rm Tl}$. It means that the approximation, $(E_i - E_n)\approx 2mc^2$, used previously overestimates the effect. In other words, negative energy states for which $(E_i - E_n) \gg 2mc^2$ also contribute to the considered PNC effect.
More specifically, according to our analysis within the PT2 approach, negative energy states $|n\rangle$ whose energy gap between the $E_n$ energy and the energy of the first occupied positive-energy state are smaller than $3mc^2$ give about 46\% of the total negative energy states contribution; negative energy states with the energy gap smaller than $6mc^2$ contribute 72\%; negative energy states with the gap smaller than $13mc^2$ contribute 91\%. The remaining part of the effect (9\%) is due to states with higher energy gap.
Such behavior can be explained by the localization of the PNC operator (\ref{NSD-PNC-rel}) inside the nucleus.

Substituting an estimation $g(^{205}{\rm Tl})\approx0.5$~\cite{Flambaum:84,GFreview} to the  final value of $J^{(1),\mathrm{PNC}}$ in terms of $g(^{205}{\rm Tl})$ one obtains $|J^{(1),\mathrm{PNC}}|\approx$ 1.5  mHz.

\section{Considerations for an experiment with TlF}
\label{Sec:Considerations_TlF}

Let us estimate the PNC signal for the apparatus and the beam used in the CeNTREX experiment \cite{CeNTREX21}. The length of the working region there is $L=2.5$\,m and the averaged beam velocity is $\langle v_z\rangle=184$\,m/s. This gives us an interaction time $T=14$\,ms. The coefficient in the Zeeman term \eqref{Zeeman_b} is
\begin{align}\label{gamma_num} 
    \left(\gamma_I -\gamma_S \right) &\approx 2\pi\cdot (-1.5\, \mathrm{kHz/G})\,.
\end{align}
This means that to have $ \Omega_{\rm M1} T =\tfrac{\pi}{2}$, we need ${\cal B}_{1}\approx 24$\,mG.

 To estimate the signal for $\tilde{J}=0$ in the presence of the static and oscillating  electric fields with amplitudes $\mathcal{E}_0 \approx 3000$ V/cm and $\mathcal{E}_1 = 1000$ V/cm we use the value $d\langle \lambda_z \rangle / d\mathcal{E}|_{\mathcal{E}_0}\mathcal{E}_1=0.079$ calculated for this field (see Fig.\ \ref{fig:n_z}). In this case $\Omega_{\rm PNC}(\tilde{J}=0) \approx 0.37\times 10^{-3}\, \mathrm{s}^{-1}$. For the case of $\tilde{J}=1$ and the same fields $\Omega_{\rm PNC}(\tilde{J}=1) \approx -0.17\times 10^{-3}\, \mathrm{s}^{-1}$.
%
%
 
 In the CeNTREX experiment it is anticipated to detect $N_d\approx 6\times 10^8$ molecules per pulse \cite{CeNTREX21}. Now we can use Eq.~\eqref{S/N} to estimate statistical sensitivity of the proposed experiment. To reach the ratio $\mathrm{S/N}=1$ for $\tilde{J}=0$ one needs a number of pulses $N_p$ given by:
\begin{align}
    \label{S/N1}
   N_p= \frac{1}{(\Omega_{\rm PNC} T)^2 N_d} \approx 62\,\mathrm{pulses}\,,
\end{align}
or about 1.24\,s at the anticipated repetition rate of 50\,Hz. In about 3.5 hours one can accumulate $\mathrm{S/N}=100$.
We conclude that statistical sensitivity of the proposed experiment is easily sufficient for an accurate measurement of the anapole moment of the thallium nucleus.

A detailed analysis of the possible systematic effects is still to be done. However, we point out that PNC experiments use reversals of experimental parameters to isolate the effect of interest from systematics, and two primary reversals are available in this case. These are (i). Reversal of the  phase (i.e., sign flip) of the oscillating $\cal{E}$-field; and (ii). Reversal of the phase (i.e., sign flip) of the oscillating $\cal{B}$-field. With only these two reversals, a possible systematic effect of concern is the following. Applying the oscillating electric field will lead to correlated charging currents and likely a component of magnetic field out of phase with the electric field. If this charging-current $\cal{B}$-field has a component along $\bf{z}$, it will add to/subtract from the deliberately applied oscillating $\cal{B}$-field and hence change the magnitude of $\Omega_{\rm M1}$ when the phase of the $\cal{E}$-field is reversed. This will exactly mimic the signature of the PNC signal. 
To combat this effect, we can make measurements not only on the rotational state $\tilde{J}=0$, but also on $\tilde{J}=1$. In these two cases, the M1 transition matrix elements are nearly the same, while the PNC signal is of opposite sign due to the opposite direction of the polarization of these states in the electric field (Fig.\,\ref{fig:n_z}) as discussed above.
Hence a properly weighted difference of the results from the two rotational states will cancel the charging-current effect but preserve the PNC effect. 

An accurate determination of the coupling constant $g_{\rm{Tl}}$ for the two stable isotopes ($^{203,205}$Tl) will allow one to study different contributions to this constant. Within the Standard Model, this constant is determined by the strength of nucleon-nucleon parity violating weak interactions \cite{haxton2001atomic, flambaum1997anapole}, which remain poorly understood \cite{gardner2017new}. 
Two prior measurements of $g_{\rm{Tl}}$ based on detection of parity-violating optical rotation in Tl atoms \cite{vetter1995FortsonTlPNC,edwards1995BairdTlPNC} gave results consistent with zero, but with uncertainties comparable to the anticipated value $g(^{205}{\rm Tl})\approx0.5$~\cite{Flambaum:84,GFreview}.
More accurate measurements of nuclear anapole moment will provide new information on PNC nuclear forces \cite{haxton2002nuclear}. In addition, such experiment will be sensitive to spin-dependent interaction with bosonic dark matter, which is predicted by some models. One of them consider interaction of nucleons with a hypothetical static pseudovector cosmic field~\cite{Roberts2014CosmicPNC}.
The constraint on the proton-cosmic field interaction parameter $b^0_p$, currently obtained~\cite{Roberts2014CosmicPNC} from the limits on the $^{203,205}$Tl anapole moment~\cite{vetter1995FortsonTlPNC,edwards1995BairdTlPNC}, can be improved when the constant $g_{\rm{Tl}}$ is measured as proposed here. Another prospect is to study a possible contribution due to hypothetical light vector bosons \cite{Dzuba2017Anapole}. The nuclear spin independent parity-violation effects due to such beyond-the-standard-model bosons were recently constrained in atomic parity-violation measurements in a chain of Yb isotopes \cite{AntypasNP2019}. 

\section{Conclusions}
Indirect electron-mediated interactions between two nuclear spins in a diatomic molecule provide an attractive route towards measurement of long sought-after molecular PNC. The technique is especially promising since, as demonstrated in the present work, (i) the PNC effect is predicted to be large enough to be measured in an ongoing experiment (CeNTREX) and (ii) it can be reliably calculated. This opens a clear path to measuring the nuclear anapole moments of the two stable isotopes, $^{205}$Tl and $^{203}$Tl, a result that would be of high importance for understanding weak interactions within nuclei \cite{haxton2001atomic,flambaum1997anapole} and search for additional vector bosons \cite{Dzuba2017Anapole}, as well as possible cosmic fields that may be related to the ``dark sector'' \cite{Roberts2014CosmicPNC}.

\acknowledgements

Electronic structure calculations have been carried out using computing resources of the federal collective usage center Complex for Simulation and Data Processing for Mega-science Facilities at National Research Centre ``Kurchatov Institute'', http://ckp.nrcki.ru/, and partly using the computing resources of the quantum chemistry laboratory.

The authors are grateful to Victor Ezhov and Vladimir Ryabov for useful comments about the possible experimental setup, and to Olivier Grasdijk and Oskari Timgren for helpful simulations of toy models. This work was supported in part by the Deutsche Forschungsgemeinschaft (DFG – German Research Foundation) Project ID 390831469:  EXC 2118 (PRISMA+ Cluster of Excellence) and Project ID 423116110; and in part by Argonne National Laboratory.
L.V.S.\ acknowledges support by the Russian Science Foundation under Grant No. 19-72-10019 for calculations of $J$-coupling tensor components  performed at NRC ``Kurchatov Institute'' -- PNPI and also acknowledges the support of the foundation for the advancement of theoretical physics and mathematics ``BASIS'' grant according to Project No. 21-1-2-47-1 for the selection rules analysis and calculation of electric field dependence of PNC matrix elements performed at SPbU. 

This manuscript is based on work completed prior to February of 2022.

\appendix
\section{Vanishing electric dipole matrix element between $\tilde{S}_0$ and $\tilde{T}_0$ states}
\label{AppNoE1}
Below we show that there is no coupling by an additional $\mathcal{E}$-field of mixed-parity singlet $|\mathrm{\tilde{S}_0}\rangle$ and triplet $|\mathrm{\tilde{T}_0}\rangle$ states in the absence of PNC effect. For this, let us consider eigenfunctions $|F,p,M_F\rangle$ of the field-free Hamiltonian 
\begin{equation}
    H_{0} =B\mathbf{J}^2 + c_1\mathbf{I}\cdot\mathbf{J}+c_2\mathbf{S}\cdot\mathbf{J}+c_3 T^{(2)}(\mathbf{C})\cdot T^{(2)}(\mathbf{I},\mathbf{S})+c_4\mathbf{I}\cdot\mathbf{S},
    \label{Hzero}
\end{equation}
which conserves quantum numbers $F$, $p$ and $M_F$, where $F$ is the value of the total angular momentum $\mathbf{F}=\mathbf{J}+\mathbf{I}+\mathbf{S}$, $M_F$ is the projection of $\mathbf{F}$ on the lab axis and $p$ is the spatial parity. We are interested in wavefunctions with $M_F=0$. For the considered case of $I=1/2$ and $S=1/2$ they can be written in the basis of functions with definite $F$ and $J$ quantum numbers as well as the total spin of Tl and F nuclei designated as $R_{IS}$:
\begin{align}
  |F,p,M_F=0\rangle &= e |F,J=F+1,R_{IS}=1,M_F=0\rangle \nonumber\\  
  &+f|F,J=F-1,R_{IS}=1,M_F=0\rangle \nonumber\\
  &+g|F,J=F,R_{IS}=1,M_F=0\rangle \nonumber\\ 
  &+h|F,J=F,R_{IS}=0,M_F=0\rangle. 
  \label{Exp2}  
\end{align}
Coefficients $e,f,g,h$ depend on the parameters of $H_0$. However, for a given parity $p$, only one of the pairs of coefficients $e,f$ or $g,h$ is nonzero, i.e. the are two classes of eigenfunctions of $H_0$ with opposite parities. Each basis function in Eq.~(\ref{Exp2}) can be further expressed in terms of the following uncoupled spin-rotational functions
\begin{align}
\label{Eq:S0}
    |J,S_0\rangle &= \frac{1}{\sqrt{2}}|J,M_J=0\rangle 
\left(|\uparrow_I\downarrow_S\rangle - |\downarrow_I\uparrow_S\rangle\right), \\
\label{Eq:T0}
|J,T_0\rangle &= \frac{1}{\sqrt{2}}|J,M_J=0\rangle 
\left(|\uparrow_I\downarrow_S\rangle + |\downarrow_I\uparrow_S\rangle\right), \\
\label{Eq:T+1}
|J,T_{-1}\rangle &= |J, M_J=1\rangle|\downarrow_I,\downarrow_S\rangle, \\
\label{Eq:T-1}
|J,T_{1}\rangle &= |J, M_J=-1\rangle|\uparrow_I,\uparrow_S\rangle
\end{align}
using standard angular momentum algebra~\cite{Varshalovich:1988}.
By applying this expansion we arrive at the following eigenfunctions of $H_0$ with definite parities:
\begin{align}
 |\Phi_{s}\rangle &= |F,p=(-1)^F,M_F=0\rangle  
 \label{PhiS} \\
 &=\alpha_J|J,S_0\rangle\ + \beta_J(|J,T_{1}\rangle - |J,T_{-1}\rangle), J=F \nonumber
\end{align}
\begin{align}
 |\Phi_{t}\rangle &= |F,p=(-1)^{F+1},M_F=0\rangle  
 \label{PhiT}  \\
 &=\sum_{J=F\pm1}
 \gamma_J|J,T_0\rangle\ + \delta_J(|J,T_{1}\rangle + |J,T_{-1}\rangle), \nonumber
\end{align}
where coefficients $\alpha_J$, $\beta_J$, $\gamma_J$, $\delta_J$ are determined by parameters of the Hamiltonian $H_0$.

Now let us consider the case with an applied uniform external electric field along the lab axis, i.e.
\begin{equation}
    H_{stat.} = H_0 + d\mathcal{E}_0 \cos\theta,
\end{equation}
where $\theta$ is the angle between molecular and lab axes.
In the present case $M_F$ is still a good quantum number while $F$ and $p$ are not. Eigenfunctions of this Hamiltonian in a given basis set can be obtained by its diagonalization. Let us calculate the following matrix element of interest in the basis of eigenfunctions of the unperturbed Hamiltonian $H_0$ considered above:
\begin{widetext}
\begin{multline}
\langle \Phi_{s}| H_{stat.} | \Phi'_{t}\rangle=
 \langle \Phi_{s}| d\mathcal{E}_0 \cos\theta | \Phi'_{t}\rangle 
 =
 \langle F,p=(-1)^F,M_F=0 | d\mathcal{E}_0 \cos\theta|  F',p'=(-1)^{F'-1},M_F=0 \rangle  
 \\
 =
 \sum_{J'=F'\pm1} \beta_J\delta_{J'}
 \left(\langle J=F,T_{1}| d\mathcal{E}_0\cos\theta| J',T_{1} \rangle - \langle J=F,T_{-1}| d\mathcal{E}_0\cos\theta| J',T_{-1} 
 \rangle\right) 
 =0.
\label{ZeroME} 
\end{multline}
\end{widetext}
In the last equality we have used Eqs.~(\ref{Eq:T+1}) and (\ref{Eq:T-1}) and the property $\langle J,M_J| \cos\theta|J',M_J\rangle=\langle J,-M_J| \cos\theta| J',-M_J\rangle$. Note also, that these matrix elements are nonzero only for $J=J'\pm1$, but in any case they either cancel each other out or both vanish in Eq.~(\ref{ZeroME}). Thus, the matrix of the Hamiltonian $H_{stat.}$ in the considered  basis set is quasi-diagonal and after its diagonalization the resulting eigenfunctions will not have a mixture of $\Phi_{s}$ and $\Phi'_{t}$ functions for any external electric field $\mathcal{E}_0$, i.e. resulting electric field-dependent functions $\tilde{\Phi}_{s}$ will be linear combinations of only $\{\Phi_{s}\}$ functions (\ref{PhiS}) with various $F$, while electric field-dependent functions $\tilde{\Phi}_{t}$ will be linear combinations of only $\{\Phi_{t}\}$~(\ref{PhiT}). According to Eq.~(\ref{ZeroME}) there would be no coupling by an additional $\mathcal{E}$-field of mixed-parity singlet $|\mathrm{\tilde{S}_0}\rangle$ and triplet $|\mathrm{\tilde{T}_0}\rangle$ states. Note, that this result also holds for the more general case when additional parity-conserving interactions inside the molecule (e.g. non-adiabatic ones) are included in the Hamiltonian. This follows from the symmetry arguments. In the static uniform external electric field the system has group of symmetry $C_{\infty v}$ with the axis of symmetry along the lab axis. Functions $\tilde{\Phi}_{s}$ and $\tilde{\Phi}_{t}$ transform according to different irreducible representations $\Sigma^+$ and $\Sigma^-$ of this group, respectively and cannot be coupled by the operator $\sim z$, transforming according to the $\Sigma^+$ irreducible representation.


%

\end{document}